\documentclass{article}

\usepackage{color}

\newtheorem{definicao}{Definition}
\newtheorem{teorema}{Theorem}

\begin{document}



\title{Finite enumerable but undecidable collections}


\author{Adonai S.
Sant'Anna\thanks{Research partially supported by CAPES (Brazilian
government agency).}}

\date{Department of Philosophy\\University of South Carolina\\Columbia, SC, 29208, USA\\
adonai{@}ufpr.br}

\maketitle


\begin{abstract}

It is well known that in Zermelo-Fraenkel (ZF) set theory any
finite set is decidable. In this paper we discuss an extension of
ZF where this result is no longer valid. Such an extension is
quasi-set theory and it has its origin on problems motivated by
quantum mechanics.

\end{abstract}


\section{Introduction}

In 1982 Richard  P. Feynman \cite{Feynman-82} proved that a
quantum system of $n$ particles cannot be simulated by an ordinary
computer without an exponential slowdown in the efficiency of the
simulation. On the other hand, a classical system with $n$
particles can be simulated with a polynomial slowdown. This was
the starting point of a new field of scientific knowledge known
today as quantum computation. For a brief review on this and
further technical details see \cite{Hirvensalo-04}.

In this paper we propose another kind of computation also inspired
on quantum phenomena. Although this new computation presents a lot
of disadvantages from the computational point of view, it may
bring some light to a better understanding of the computational
aspects of the quantum world.

It is well known that in quantum mechanics (QM) elementary
particles may be considered as non-individuals in a sense. Quantum
particles that share the same set of state-independent (intrinsic)
properties may be {\em indistinguishable\/}. Although classical
particles can share all their intrinsic properties, we are able to
follow their trajectories, at least in principle. That allows us
to identify particles. In quantum physics this is not possible,
i.e., it is not possible, {\em a priori\/}, to keep track of
individual particles in order to distinguish among them. In other
words, it is not possible to label quantum particles by their
trajectories. And this non-individuality plays a very important
role in quantum mechanics \cite{Sakurai-94}. For a philosophical
discussion on the problems raised by non-individuality see, for
example, the references in \cite{French-04}.

On the possibility that collections of such indistinguishable
entities should not be considered as sets in the usual sense, Yu.
Manin \cite{Manin-76} proposed the search for axioms which should
allow to deal with indiscernible objects. As he said,

\begin{quote}
I would like to point out that it [standard set theory] is rather
an extrapolation of common-place physics, where we can distinguish
things, count them, put them in some order, etc. New quantum
physics has shown us models of entities with quite different
behavior. Even {\em sets\/} of photons in a looking-glass box, or
of electrons in a nickel piece are much less Cantorian than the
{\em sets\/} of grains of sand.
\end{quote}

It is important to settle that `indistinguishable' objects are
objects that share their properties, while `identical' objects
means `the very same object'. One manner to cope with the problem
of non-individuality in quantum physics is by means of quasi-set
theory \cite{Krause-92,Krause-99,Sant'Anna-00}, which is an
extension of Zermelo-Fraenkel set theory that allows us to talk
about certain indistinguishable objects that are not necessarily
identical. Actually, in some cases there is no sense in saying
that two objects are either identical or different. In quasi-set
theory identity does not apply to all objects. In other words,
there are some situations in quasi-set theory where the sequence
of symbols $x = y$ is not a well-formed formula, i.e., it is
meaningless. A weaker equivalence relation called
``indistinguishability'' is an extension of identity in the sense
that it allows the existence of {\em two\/} objects that are
indistinguishable. In standard mathematics, there is no sense in
saying that two objects are identical. If $x = y$, then we are
talking about one single object with two labels, namely, $x$ and
$y$.

Some applications of quasi-set theory on the foundations of
quantum physics have already been done ({\em op. cit.\/}). But in
this paper we intend to explore some computational properties of
quasi-sets, although the quantum perspective is still present. We
intend to prove that there exist finite collections of objects in
quasi-set theory which are enumerable but undecidable. We recall
that a collection is said to be enumerable if there is an
algorithm that prints all elements of $x$ and only them. Besides,
a collection is said to be decidable if there exists an algorithm
that determines whether an arbitrary object belongs to $x$ or not.
Otherwise, the collection is undecidable. In this sense we are
extending the definition given in some textbooks like
\cite{Shen-03}.

Some related topic are discussed at the end of the paper.

\section{Quasi-sets}

This section is strongly based on other works
\cite{Krause-92,Krause-99,Sant'Anna-00}. I use standard logical
notation for first-order theories without identity
\cite{Mendelson-97}.

It is important to remark that, in contrast to the notions of set
and quasi-set, the term ``collection'' has an intuitive meaning in
this paper.

Quasi-set theory ${\cal Q}$ is based on Zermelo-Fraenkel-like
axioms and allows the presence of two sorts of atoms ({\it
Urelemente\/}), termed $m$-atoms (micro-atoms) and $M$-atoms
(macro-atoms). Concerning the $m$-atoms, a weaker `relation of
indistinguishability' (denoted by the symbol $\equiv$), is used
instead of identity, and it is postulated that $\equiv$ has the
properties of an equivalence relation. The predicate of equality
cannot be applied to the $m$-atoms, since no expression of the
form $x = y$ is a formula if $x$ or $y$ denote $m$-atoms. Hence,
there is a precise sense in saying that $m$-atoms can be
indistinguishable without being identical.

The universe of ${\cal Q}$ is composed by $m$-atoms, $M$-atoms and
{\it quasi-sets\/}. The axiomatization is adapted from that of ZFU
(Zermelo-Fraenkel with {\it Urelemente\/}), and when we restrict
the theory to the case which does not consider $m$-atoms,
quasi-set theory is essentially equivalent to ZFU, and the
corresponding quasi-sets can then be termed `sets' (similarly, if
also the $M$-atoms are ruled out, the theory collapses into ZFC).
The $M$-atoms play the same role of the {\it Urelemente\/} in ZFU.

In all that follows, $\exists_Q$ and $\forall_Q$ are the
quantifiers relativized to quasi-sets. That is, $Q(x)$ reads as
`$x$ is a quasi-set'.

In order to preserve the concept of identity for the
`well-behaved' objects, an {\it Extensional Equality\/} is defined
for those entities which are not $m$-atoms on the following
grounds: for all $x$ and $y$, if they are not $m$-atoms, then $$x
=_{E} y := \forall z ( z \in x \Leftrightarrow z \in y ) \vee
(M(x) \wedge M(y) \wedge x \equiv y).$$

It is possible to prove that $=_{E}$ has all the properties of
classical identity in a first order theory and so these properties
hold regarding $M$-atoms and `sets'. This happens because one of
the axioms of quasi-set theory says that the axiom of
substitutivity of standard identity holds only for extensional
equality. Concerning the more general relationship of
indistinguishability nothing else is said. In symbols, the first
axioms of ${\cal Q}$ are:

\begin{itemize}

\item $\forall x (x \equiv x)$,

\item $\forall x \forall y (x \equiv y \Rightarrow y \equiv x)$,
and

\item $\forall x \forall y \forall z (x \equiv y \wedge y \equiv z
\Rightarrow x \equiv z)$.

\end{itemize}

And the fourth axiom says that

\begin{itemize}

\item $\forall x \forall y (x =_{E} y \Rightarrow (A(x,x)
\Rightarrow A(x,y)))$, with the usual syntactic restrictions on
the occurrences of variables in the formula $A$.

\end{itemize}

In this text, all references to `$=$' (in quasi-set theory) stand
for `$=_E$', and similarly `$\leq$' and `$\geq$' stand,
respectively, for `$\leq_E$' and `$\geq_E$'. Among the specific
axioms of ${\cal Q}$, few of them deserve a more detailed
explanation. The other axioms are adapted from ZFU.

For instance, to form certain elementary quasi-sets, such as those
containing `two' objects, we cannot use something like the usual
`pair axiom', since its standard formulation assumes identity; we
use the weak relation of indistinguishability instead:

\begin{quote}

The `Weak-Pair' Axiom - For all $x$ and $y$, there exists a
quasi-set whose elements are the indistinguishable objects from
either $x$ or $y$. In symbols,

$$\forall x \forall y \exists_{Q} z \forall t (t \in z
\Leftrightarrow t \equiv x \vee t \equiv y).$$

\end{quote}

Such a quasi-set is denoted by $[x, y]$ and, when $x \equiv y$, we
have $[x]$, by definition. We remark that this quasi-set {\it
cannot\/} be regarded as the `singleton' of $x$, since its
elements are {\it all\/} the objects indistinguishable from $x$,
so its `cardinality' (see below) may be greater than $1$. A
concept of {\it strong singleton\/}, which plays a crucial role in
the applications of quasi-set theory, may be defined.

In ${\cal Q}$ we also assume a Separation Schema, which
intuitively says that from a quasi-set $x$ and a formula
$\alpha(t)$, we obtain a sub-quasi-set of $x$ denoted by $$[t\in x
: \alpha(t)].$$

We use the standard notation with `$\{$' and `$\}$' instead of
`$[$' and `$]$' only in the case where the quasi-set is a {\it
set\/}.

It is intuitive that the concept of {\it function\/} cannot also
be defined in the standard way, so a weaker concept of {\it
quasi-function\/} was introduced, which maps collections of
indistinguishable objects into collections of indistinguishable
objects; when there are no $m$-atoms involved, the concept is
reduced to that of function as usually understood. Relations (or
{\em quasi-relations\/}), however, can be defined in the usual
way, although no order relation can be defined on a quasi-set of
indistinguishable $m$-atoms, since partial and total orders
require antisymmetry, which cannot be stated without identity.
Asymmetry also cannot be supposed, for if $x \equiv y$, then for
every relation $R$ such that $\langle x, y \rangle \in R$, it
follows that $\langle x, y \rangle =_{E} [[x]] =_{E} \langle y, x
\rangle \in R$, by force of the axioms of ${\cal Q}$.

It is possible to define a translation from the language of ZFU
into the language of ${\cal Q}$ in such a way that we can obtain a
`copy' of ZFU in ${\cal Q}$. In this copy, all the usual
mathematical concepts (like those of cardinal, ordinal, etc.) can
be defined; the `sets' (actually, the `${\cal Q}$-sets' which are
`copies' of the ZFU-sets) turn out to be those quasi-sets whose
transitive closure (this concept is like the usual one) does not
contain $m$-atoms.

Although some authors like Weyl \cite{Weyl-49} sustain that
(concerning cardinals and ordinals) ``the concept of ordinal is
the primary one'', quantum mechanics seems to present strong
arguments for questioning this thesis, and the idea of presenting
collections which have a cardinal but not an ordinal is one of the
most basic and important assumptions of quasi-set theory.

The concept of {\it quasi-cardinal\/} is taken as primitive in
${\cal Q}$, subject to certain axioms that permit us to operate
with quasi-cardinals in a similar way to that of cardinals in
standard set theories. Among the axioms for quasi-cardinality, we
mention those below, but first we recall that in ${\cal Q}$,
$qc(x)$ stands for the `quasi-cardinal' of the quasi-set $x$,
while $Z(x)$ says that $x$ is a {\it set\/} (in ${\cal Q}$).
Furthermore, $Cd(x)$ and $card(x)$ mean `$x$ is a cardinal' and
`the cardinal of $x$', respectively, defined as usual in the
`copy' of ZFU.

\begin{quote}

Quasi-cardinality - Every quasi-set has an unique quasi-cardinal
which is a cardinal (as defined in the `ZFU-part' of the theory)
and, if the quasi-set is in particular a set, then this
quasi-cardinal is its cardinal {\em stricto sensu}:

$$\forall_{Q} x \exists_{Q} ! y (Cd(y) \wedge y =_{E} qc(x) \wedge
(Z(x) \Rightarrow y =_{E} card(x))).$$

\end{quote}

From the fact that $\emptyset$ is a set, it follows that its
quasi-cardinality is 0 (zero).

${\cal Q}$ still encompasses an axiom which says that if the
quasi-cardinal of a quasi-set $x$ is $\alpha$, then for every
quasi-cardinal $\beta \leq \alpha$, there is a sub-quasi-set of
$x$ whose quasi-cardinal is $\beta$, where the concept of {\it
sub-quasi-set\/} is like the usual one. In symbols,

\begin{quote}

The quasi-cardinals of sub-quasi-sets - $$\forall_{Q} x (qc(x)
=_{E} \alpha \Rightarrow \forall \beta (\beta \leq_{E} \alpha
\Rightarrow \exists_{Q} y (y \subseteq x \wedge qc(y) =_{E}
\beta)).$$

\end{quote}

Another axiom states that

\begin{quote}

The quasi-cardinal of the power quasi-set -
$$\forall_{Q} x (qc({\cal P}(x)) =_{E} 2^{qc(x)}).$$

\noindent where $2^{qc(x)}$ has its usual meaning.

\end{quote}

These last axioms allow us to talk about the quantity of elements
of a quasi-set, although we cannot count its elements in many
situations.

As remarked above, in ${\cal Q}$ there may exist quasi-sets whose
elements are $m$-atoms only, called `pure' quasi-sets.
Furthermore, it may be the case that the $m$-atoms of a pure
quasi-set $x$ are indistinguishable from one another. In this
case, the axiomatization provides the grounds for saying that
nothing in the theory can distinguish among the elements of $x$.
But, in this case, one could ask what it is that sustains the idea
that there is more than one entity in $x$. The answer is obtained
through the above mentioned axioms (among others, of course).
Since the quasi-cardinal of the power quasi-set of $x$ has
quasi-cardinal $2^{qc(x)}$, then if $qc(x) = \alpha$, for every
quasi-cardinal $\beta \leq \alpha$ there exists a sub-quasi-set $y
\subseteq x$ such that $qc(y) = \beta$, according to the axiom
about the quasi-cardinality of the sub-quasi-sets. Thus, if $qc(x)
= \alpha \not= 0$, the axiomatization does not forbid the
existence of $\alpha$ sub-quasi-sets of $x$ which can be regarded
as `singletons'.

Of course the theory cannot prove that these `unitary'
sub-quasi-sets (supposing now that $qc(x) \geq 2$) are distinct,
since we have no way of `identifying' their elements, but
quasi-set theory is compatible with this idea. In other words, it
is consistent with ${\cal Q}$ to advocate that $x$ has $\alpha$
elements, which may be regarded as absolutely indistinguishable
objects. Since the elements of $x$ may share the relation
$\equiv$, they may be further understood as belonging to the same
`equivalence class' but in such a way that we cannot assert either
that they are identical or that they are distinct from one
another.

The collections $x$ and $y$ are defined as {\it similar\/}
quasi-sets (in symbols, $Sim(x,y)$) if the elements of one of them
are indistinguishable from the elements of the other one, that is,
$Sim(x,y)$ if and only if $\forall z \forall t (z \in x \wedge t
\in y \Rightarrow z \equiv t)$. Furthermore, $x$ and $y$ are {\it
Q-Similar\/} ($QSim(x,y)$) if and only if they are similar and
have the same quasi-cardinality. Then, since the quotient
quasi-set $x/_{\equiv}$ may be regarded as a collection of
equivalence classes of indistinguishable objects, then the `weak'
axiom of extensionality is:

\begin{quote}

Weak Extensionality -
\begin{eqnarray}
\forall_{Q} x \forall_{Q} y (\forall z (z \in x/_{\equiv}
\Rightarrow \exists t (t \in y/_{\equiv} \wedge \, QSim(z,t))
\wedge \forall t(t \in
y/_{\equiv} \Rightarrow\nonumber\\
\exists z (z \in  x/_{\equiv} \wedge \, QSim(t,z)))) \Rightarrow x
\equiv y)\nonumber
\end{eqnarray}

\end{quote}

In other words, this axiom says that those quasi-sets that have
the same quantity of elements of the same sort (in the sense that
they belong to the same equivalence class of indistinguishable
objects) are indistinguishable.

\begin{definicao}
A {\em strong singleton\/} of $x$ is a quasi-set $x'$ which
satisfies the following property:
$$x' \subseteq [x] \wedge qc(x') =_{E} 1$$
\end{definicao}

\begin{definicao}
A {\em $n$-singleton\/} of $x$ is a quasi-set $[x]_n$ which
satisfies the following property:
$$[x]_n \subseteq [x] \wedge qc([x]_n) =_{E} n$$
\end{definicao}

\section{Finite enumerable but undecidable quasi-sets}

This section introduces the main contributions of this paper. The
next definition is crucial for our purposes. It is important to
recall that if $x$ is a term, then $x'$ is a strong singleton
whose only element is indistinguishable from $x$.

\begin{definicao}
If $x$ is a quasi-set and $y$ is indistinguishable from a given
element $z$ that belongs to $x$, then $$x\ominus y'=_E x-z',
\mbox{ where } z'\subseteq x.$$
\end{definicao}

We call $\ominus$ the {\em strong difference\/} between
quasi-sets. This operation allows us to drop one of the elements
of $x$. So, if $qc(x) = n$ and $n$ is a natural number, then
$qc(x\ominus y') = n-1$.

Now we introduce an algorithm which allows us to prove that every
finite quasi-set is enumerable. We recall that a set $x$ is
enumerable if there is an algorithm that prints all elements of
$x$ and only them. We first prove the most interesting case where
all elements of the given quasi-set are micro-atoms of the same
type (i.e., indistinguishable). Other cases may be proved by
similar arguments.

\begin{teorema}
If $[x]_n$ is a finite $n$-singleton, then it is enumerable.
\end{teorema}

{\bf Proof:} Consider the following algorithm,

\begin{description}

\item[1.] INPUT $[ x ]_n$

\item[2.] DO $y := [x]_n$

\item[3.] DO $[ x ]_{n-1} := [ x ]_n \ominus x'$

\item[4.] PRINT $y-[x]_{n-1}$

\item[5.] DO $n := n-1$

\item[6.] IF $[ x ]_n =_E \emptyset$ THEN GO TO {\bf 8}

\item[7.] GO TO {\bf 2}

\item[8.] END

\end{description}

In the first step, we introduce a finite $n$-singleton $[ x ]_n$,
i.e., a pure quasi-set with a finite quasi-cardinality (a finite
number of elements) where all its elements are indistinguishable
objects of the same kind. In the second step we attribute $[x]_n$
to $y$, which means that $y$ and $[x]_n$ are extensionally
identical. Next, we perform the strong difference in order to drop
one of the elements from the quasi-set $[ x ]_n$ and attribute
this new collection to $[x]_{n-1}$. Then we print the element that
was subtracted from $[x]_n$. We repeat this process until $[x]_n$
gets empty.\colorbox{black}{\textcolor{white}{ }}

\vskip3mm

So, we printed all the elements of the original $[x]_n$, which
means that $[x]_n$ is enumerable.

Now we will prove that even a finite enumerable quasi-set may be
undecidable. We recall that a collection $x$ is decidable if there
exists an algorithm that determines whether an arbitrary object
belongs or not to $x$. Otherwise, $x$ is said undecidable.

\begin{teorema}
If $[x]_n$ is a non-empty finite $n$-singleton, then it is
undecidable.
\end{teorema}

{\bf Proof:} If $y\equiv x$, then there is no way to know if $y$
belongs to $[x]_n$ or not. Actually we cannot even know if $x\in
[x]_n$, although we always know that $x\in [x]$. This happens
because in quasi-set theory it is legitimate the existence of many
indistinguishable objects. So, $[x]_n$ is
undecidable.\colorbox{black}{\textcolor{white}{ }}\\

\section{Final Remarks}

There are many results concerning undecidability in mathematics
and even in physics. See, for example, \cite{daCosta-91}, where
the authors derive a general undecidability and incompleteness
result for elementary functions within Zermelo-Fraenkel set theory
(with the axiom of choice), and apply it to some important
problems in Hamiltonian mechanics and dynamical systems.

But all results on undecidability, as far as we know, refer to
infinite sets or collections. In this paper we believe that we are
presenting for the first time an example of a finite system that
is undecidable. This is due to the fact that although in standard
mathematics the membership relationship seems to present some
tricky features when we talk about infinite collections, in
quasi-set theory there is another tricky relationship, namely,
indistinguishability.

We do not know if the weak singleton $[x]$ is enumerable or not
(open problem). But by using similar arguments we can easily prove
that $[x]$ is undecidable, if it is not empty.

\section{Acknowledgements}

This paper was quite improved thanks to some discussions with
Ot\'avio Bueno, Newton C. A. da Costa, and D\'ecio Krause.

I would like to thank Ot\'avio Bueno and Davis Baird for their
hospitality during my stay at the Department of Philosophy of the
University of South Carolina as a Visiting Scholar.


\begin{thebibliography}{99}

\bibitem{daCosta-91} da Costa, N. C. A. and F. A. Doria (1991),
`Undecidability and incompleteness in classical mechanics', {\em
International Journal of Theoretical Physics\/} {\bf 30}
1041-1074.

\bibitem{Feynman-82} Feynman, R. P. (1982), `Simulating physics
with computers', {\em International Journal of Theoretical
Physics\/} {\bf 21} 467-488.

\bibitem{French-04} French, S. (2004), `Identity and individuality in quantum
theory', The Stanford Encyclopedia of Philosophy, Edward N. Zalta
(ed.), URL = http://plato.stanford.edu/entries/qt-idind/

\bibitem{Hirvensalo-04} Hirvensalo, M. (2004), {\em Quantum Computing\/}
Springer, Berlin.

\bibitem{Krause-92} Krause, D. (1992), `On a quasi-set theory' {\em Notre
Dame Journal of Formal Logic\/} {\bf 33} 402-411.

\bibitem{Krause-99} Krause, D., A. S. Sant'Anna and A. G. Volkov (1999),
`Quasi-set theory for bosons and fermions: quantum distributions',
{\em Found. Phys. Lett.\/}, {\bf 12} 51-66.

\bibitem{Manin-76} Manin, Yu. I. (1976), `Problems of Present Day
Mathematics I: Foundations', in Browder, F. E. (ed.), {\em
Mathematical problems arising from Hilbert problems\/},
Proceedings of Symposia in Pure Mathematics XXVIII, Providence,
AMS, 36-36.

\bibitem{Mendelson-97} Mendelson, E. (1997), {\em Introduction to
Mathematical Logic\/} (Chapman \& Hall, London).

\bibitem{Sakurai-94} Sakurai, J. J. (1994), {\em Modern Quantum Mechanics\/}
(Addison-Wesley, Reading).

\bibitem{Sant'Anna-00} Sant'Anna, A. S. and A. M. S. Santos (2000),
`Quasi-set-theoretical foundations of statistical mechanics: a
research program', {\em Found. Phys.\/}, {\bf 30} 101-120.

\bibitem{Shen-03} Shen, A., Vereshchagin, N. K.: {\em Computable
functions\/}, American Mathematical Society, 2003.

\bibitem{Weyl-49} Weyl, H., {\it Philosophy of Mathematics and
Natural Science\/}, (Princeton Un. Press, Princeton, 1949).

\end{thebibliography}

\end{document}